\documentclass[english,12pt]{article}

\usepackage{authblk}
\usepackage[T1]{fontenc}
\usepackage[utf8]{inputenc}
\usepackage{geometry}
\usepackage{babel}
\usepackage{amsthm}
\usepackage{amsmath}
\usepackage{amssymb}
\usepackage{mathrsfs}
\usepackage{mathtools}
\usepackage{setspace}
\usepackage{bbm}
\usepackage{babel}
\usepackage{natbib}
\usepackage{appendix}  
\usepackage{pdfpages}
\usepackage[open,openlevel=1]{bookmark}
\usepackage{xcolor}
\hypersetup{
    colorlinks,
    linkcolor={red!50!black},
    citecolor={blue!50!black},
    urlcolor={blue!80!black}
}
\usepackage{multirow}
\usepackage{tikz}

\usepackage{thmtools, thm-restate}  
\declaretheorem{theorem}  

\declaretheoremstyle[
spaceabove=6pt, spacebelow=6pt,
headfont=\normalfont\bfseries,
notefont=\mdseries, notebraces={(}{)},
bodyfont=\normalfont,
postheadspace=1em,
qed=$\blacksquare$
]{examplestyle}

\renewcommand\thmcontinues[1]{continued}

\newcommand{\real}{\mathbb{R}}

\newcommand{\norm}[1]{\left\Vert #1\right\Vert }
\newcommand{\indicator}{\mathbbm{1}}
\newcommand{\pd}[2]{\frac{\partial#1}{\partial#2}}

\newcommand{\ud}{\mathrm{d}}

\newcommand{\prob}{\mathbb{P}}
\newcommand{\expt}{\mathbb{E}}

\newcounter{steps}

\DeclareMathOperator*{\argmin}{arg\,min}

\usepackage{enumitem}
\setenumerate[1]{label=(\roman*)}
\setenumerate[2]{label=\normalfont{(\alph*)}}

\title{A General Method for Demand Inversion}
\author{Lixiong Li \footnote{Email: \href{mailto:lixiong@psu.edu}{lixiong@psu.edu}}}
\affil{Pennsylvania State University}
\date{February 26, 2018}

\begin{document}
\maketitle
\abstract{This paper describes a numerical method to solve for mean product
  qualities which equates the real market share to the market share predicted by
  a discrete choice model. The method covers a general class of discrete choice
  model, including the pure characteristics model in \cite{berry_pure_2007} and
  the random coefficient logit model in \cite{berry_automobile_1995} (hereafter
  BLP). The method transforms the original market share inversion problem to an
  unconstrained convex minimization problem, so that any convex programming
  algorithm can be used to solve the inversion. Moreover, such results also
  imply that the computational complexity of inverting a demand model should be
  no more than that of a convex programming problem. In simulation examples, I
  show the method outperforms the contraction mapping algorithm in BLP. I also
  find the method remains robust in pure characteristics models with near-zero
  market shares. }
\newpage
\section{Introduction}
When conducting structural demand estimations, one usually needs to solve the
inversion of a demand model given the real market share. The problem often takes
the form of a nonlinear equation,
\begin{equation}
 \text{find }x\text{ such that } \sigma(x) = \sigma^* \label{eq:dgosn}
\end{equation}
where $\sigma^*$ is a vector of real market share and $\sigma(x)$ is the model
predicted market share given the demand shifter (or mean product quality) $x$.
To ensure that the entire estimation procedure is robust and tractable, this
inversion problem needs to be solved both precisely and efficiently.

In this paper, I describes a method which transforms the inversion problem
\eqref{eq:dgosn} into an unconstrained convex optimization problem. This
transformation works for a general class of discrete choice models. In
particular, it covers the pure characteristics model in \cite{berry_pure_2007}
and the random coefficient logit model in \cite{berry_automobile_1995}
(henceforth BLP). As a result of the convex transformation, any convex
optimization algorithms can be used to solve the inversion problem. As the
algorithms of convex optimization have been extensively studied and their robust
implementation are now widely available, the method here is able to solve the
inversion problem in \eqref{eq:dgosn} efficiently and robustly in both small and
large scale problems.

When the demand function $\sigma(\cdot)$ is generated by a mixed logit model,
BLP provides an inversion algorithm based on
contraction mapping. Their algorithm converges globally at a \emph{linear} rate,
i.e. there exists some $\alpha \in (0, 1)$ such that
\begin{equation}
  \label{eq:linear_contraction}
  \frac{\norm{x^{(k+1)} - x^*}}{\norm{x^{(k)} - x^*}} \le \alpha,
\end{equation}
where $x^*$ is the solution of \eqref{eq:dgosn} and $x^{(k)}$ is the best solution
of the algorithm at $k$th iteration. In practice, $\alpha$ could be very close
to $1$, which leads to a very slow convergence rate. See our simulation below
and \cite{dube_improving_2012}. In contrast, algorithms designed for
unconstrained smooth convex optimization can typically converge globally at a
\emph{superlinear} rate, i.e.
\begin{displaymath}
  \lim_{k\to \infty} \frac{\norm{x^{(k+1)} - x^*}}{\norm{x^{(k)} - x^*}} = 0
\end{displaymath}
when the Jacobian matrix of $\sigma$ is non-singular. See, for example, Theorem
5.13 in \cite{ruszczynski_nonlinear_2006}. As illustrated in simulation
experiments, such a difference in convergence speeds could result in considerable
efficiency gains in practice.

When the demand function $\sigma(\cdot)$ is generated by pure characteristics
model, \cite{berry_pure_2007} provides three algorithms to solve the inversion
problem, and propose to combine all the algorithms, as none of them works on its
own. \cite{song_estimating_2006} proposes a detailed procedure on how to combine
the Newton-Raphson method into those algorithms. Yet, none of those are fully
satisfactory. In particular, none of the existing algorithms ensures global
convergence. Also, in practice, when those methods do converges in some cases, their
convergence rate tend to be slow. The main challenge here is that the
Jacobian matrix of $\sigma$ could be near-singular or singular when the market
share of some products is very close or equal to zero. This could happen even if
each product has positive share at $\sigma^*$, since algorithms can drive $x^{(k)}$
into regions of zero market share during the iteration.

In contrast, the method here still works even if the Jacobian of $\sigma$ is
singular. This is due to the fact that some algorithms of convex optimization
could still have global convergence (though not necessarily at superlinear rate)
even if the Hessian matrix is singular. Among them, the trust region algorithms are very suitable
in those singular cases, whose convergence properties and robustness have been
extensively studied and developed since 1970's. For example, see Theorem 5.1.1
in \cite{fletcher_practical_1987} and more general results in
\cite{shultz_family_1985} and \cite{conn_trust_region_2000}.

The rest of the paper is organized as follows. Section \ref{sec:disc_choice}
presents how to transform the inversion problem into a convex minimization
problem. Section \ref{sec:simu} conducts simulation experiments on random
coefficients logit models and pure characteristics models. All proofs are
contained in the Appendix.

\section{Convex Transformation}\label{sec:disc_choice}
Consider a discrete choice model of $J$ inside products and one outside option.
Assume agent $i$'s uility of choosing product $j$ is
\begin{displaymath}
  u_{ji} \coloneqq x_j + \epsilon_{ij}
\end{displaymath}
where $x_j$ is the demand shifter (or mean utility) of product $j$, and
$\epsilon_{ij}$ is the taste shock of agent $i$. Normalize agent $i$'s utility
from consuming the outside option to be zero. Let $x\coloneqq (x_1,...,x_J)$ and
$\epsilon_i \coloneqq (\epsilon_{i1}, ..., \epsilon_{iJ})$ be vectors in
$\real^J$.

The above framework covers most discrete choice models used in empirical
applications. For example, in random coefficients logit models as in BLP, we
have $x_j = \xi_j + z_j' \beta$ and $\epsilon_{ij} = z_j' \nu_i +
\varepsilon_{ij}$ where $\xi_j$ is the unobserved product quality, $z_j$ is the
vector of both exogenous and endogenous product attributes, $\beta$ is the
parameter and $\nu_i$ is the random coefficient of consumer $i$ and
$\varepsilon_{ij}$ is the logistic shock of consumer $i$.

Let $F$ be the distribution of $\epsilon_i$. Then, the average welfare of
consumer $U(x)$ can be written as
\begin{displaymath}
  U(x) = \expt_F \max\left(0,  \max_{j=1,...,J}(x_j + \epsilon_{ij}) \right).
\end{displaymath}

Let $\sigma(x)\in\real^J$ be the market share of inside goods given $x$ and
distribution $F$ and let $\Delta^J$ be defined as the simplex
\begin{displaymath}
  \Delta^J \coloneqq \Bigg\{(s_1,...,s_J) \in \real^J: \forall j,\  s_j \ge 0\text{ and
}\sum_j s_j \le 1 \Bigg\}.
\end{displaymath}

Given $F$ and some $\sigma^* \in \Delta^J$, the inversion problem of the
discrete choice model is to find an $x^*$ such that $\sigma(x^*) = \sigma^*$
whenever such $x^*$ exists. The following theorem transforms such inversion
problem into an unconstrained convex optimization problem.

\begin{theorem}\label{thm:convex_representation}
  Given any $\sigma^* \in \Delta^J$, $\sigma(x^*) = \sigma^*$ implies 
    \begin{equation}
      \label{eq:convex_conjugate}
    x^* \in \argmin_{x} \big(U(x) - x'\sigma^*\big).
  \end{equation}
  If distribuion $F$ is non-atomic, the reverse is also true, i.e.
  condition \eqref{eq:convex_conjugate} implies $\sigma(x^*) = \sigma^*$.
\end{theorem}

Theorem \ref{thm:convex_representation} provides a necessary condition for $x^*$
to be a solution of the inversion problem. When distribution $F$ is non-atomic,
as in most empirical applications, such condition is also sufficient. Since
$U(x)$ is a convex function of $x$, solving the inversion problem should be as
easy as solving an unconstrained convex optimization problem. Moreover, when $F$
is absolutely continuous, $U(x)$ is twice differentiable so that all smooth
convex optimization algorithms can be applied to solve for $x^*$ in
\eqref{eq:convex_conjugate}.



To see the intuition behind Theorem \ref{thm:convex_representation}, let's assume
distribution $F$ is non-atomic. Then, $x^*$ is a minimizer of  $U(x) -
x'\sigma^*$ if and only if $x^*$ satisfies the first order condition,
\begin{equation}\label{eq:ooafwe}
  \pd{U(x^*)}{x} = \sigma^*
\end{equation}
One can verify that
\begin{equation}\label{eq:cnu8ya}
\forall j=1,...,J,\quad   \pd{U(x^*)}{x_j} = \prob \Bigg[ x^*_j + \epsilon_{ij} \ge  \max\Big(0,
    \max_{j=1,...,J}(x^*_j + \epsilon_{ij}) \Big)   \Bigg]  = \sigma(x^*)
\end{equation}
Combine condition \eqref{eq:ooafwe} and \eqref{eq:cnu8ya}, we know $x^*$
minimizes $U(x) - x'\sigma^*$ if and only if $\sigma(x^*) = \sigma^*$. When $F$
is atomic, condition \eqref{eq:ooafwe} no longer holds as $U(x)$ could be
nondifferentiable, but the same intuition still carries through. See Appendix
\ref{sec:proof} for a formal proof of Theorem \ref{thm:convex_representation}.

In practice, one might consider $\sigma(x) = \sigma^*$ as a general nonlinear
equation and solve it using Netwon or Quasi-Netwon methods. This approach is
known to perform well when the initial value is close enough to the true value,
but could fail to converge otherwise. To see how it's related to the method
here, note that if we apply Netwon's method to solve
\eqref{eq:convex_conjugate}, we get
\begin{equation}
  \label{eq:isawef}
  x^{(k+1)} = x^{(k)} - H_k^{-1} \Big( \pd{U(x^*)}{x}- \sigma^*\Big)
\end{equation}
where $x^{(k)}$ denotes the solution at iteration $k$ and $H_k$ stands for the
Hessian matrix of $U(x)$ at $x^{(k)}$. However, in equation \eqref{eq:cnu8ya}, we
know $\partial U(x)/\partial x = \sigma(x)$ and therefore $H_k$ equals the
Jacobian matrix of $\sigma(x)$ at $x^{(k)}$. As a result, iteration
\eqref{eq:isawef} is equivalent to
\begin{displaymath}
  x^{(k+1)} = x^{(k)} - H_k^{-1} \cdot (\sigma(x^{(k)}) - \sigma^*)
\end{displaymath}
which actually is the iteration step when we apply Newton-Raphson's method to
solve equation $\sigma(x) = \sigma^*$. However, there is a key difference
between solving the convex optimization problem in \eqref{eq:convex_conjugate}
and solving the nonlinear equation $\sigma(x) = \sigma^*$: Theorem
\ref{thm:convex_representation} enable us to utilize the objective function
$U(x) - x'\sigma^*$. Such extra information is not used in solving nonlinear
equation $\sigma(x) = \sigma^*$, but is utilized in solving its transformed
problem \eqref{eq:convex_conjugate}. Indeed, this extra information plays an
essential role in ensuring the global convergence of convex optimization
algorithms, especially when $H_k$ is singular or nearly singular. As an example
of such global convergence results, see Theorem 5.1.1 in
\cite{fletcher_practical_1987}. In the following simulation experiments, we can
see clearly how this difference results in different degrees of robustness.

\section{Simulation Experiments}\label{sec:simu}
In this section, I conduct simulation experiments on random coefficients logit models and pure
characteristics models. In each experiment, I solve the inversion problem by
minimizing the convex function in Theorem \ref{thm:convex_representation}, and
compare its robustness and speed with other methods. 

\subsection{Random Coefficients Logit Models}
I construct the following random coefficients logit model. Let $J$ denote the
number of products and $M$ denote the dimension of product attributes. In each
simulation, we draw parameter $\beta$ from uniform distribution in $[0, 1]^M$,
and draw product attributes $z_1,...,z_J$ independently from a normal distribution
$N(0, I_M)$ where $I_M$ is the $M$ dimensional identity matrix. We then draw $n
= 5000$ i.i.d. random coefficients $\nu_i$ from $N(0, I_M)$. Given $\beta$,
$\{z_j\}$ and $\{\nu_i\}$ in each simulation, consumer welfare $U(x)$ and market
share $\sigma(x)$ can be calculated as
\begin{eqnarray*}
  U(x) & = & \frac{1}{n}\sum_{i=1}^n \log\Big(1 + \sum_{q=1}^J\exp(x_{q} + z'_{q}
                    \nu_{i})\Big) + \gamma \\
  \sigma_j(x) & = & \frac{1}{n}\sum_{i=1}^n \frac{\exp(x + z_j'\nu_i)}{1 +
                    \sum_{q=1}^J\exp(x_{q} + z'_{q} \nu_{i})} \quad j = 1,...,J
\end{eqnarray*}
where $\gamma$ is Euler's constant. Let
\begin{eqnarray*}
  x^* & = & (z_1'\beta,...,z_J'\beta) \\
  \sigma^*  & = & \sigma(x^*)
\end{eqnarray*}
In the simulation experiments, I solve the inversion problem $\sigma(x) =
\sigma^*$ using the following two methods:
\begin{enumerate}
  \item \label{enu:M1}Contraction Mapping in \cite{berry_automobile_1995}.  
  \item\label{enu:M3} Trust-region algorithm applied to convex optimization
      \eqref{thm:convex_representation}.
\end{enumerate}
As both methods ensures global convergence, our focus here is to compare their
convergence rate.

I conduct $100$ simulation experiments with $J = 10$ and $M = 5$. Within each
simulation experiment, a new set of $\beta$, $\{z_j\}$ and $\{\nu_j\}$ is drawn
and each algorithm takes $x_1 = x^* + \delta$ as its initial point, where
$\delta$ is a random vector drawn in each experiment with $\norm{\delta} = 20$.
In each iteration $k$, I record the maximum norm of the error,
$\norm{\sigma(x^{(k)}) - \sigma^*}_{\infty}\coloneqq\max_j(|\sigma_j(x^{(k)}) -
\sigma^*_j|)$, where $x^{(k)}$ is the best solution up to iteration $k$.
Algorithm \ref{enu:M1} calculates $\sigma(x)$ once every iteration, and
Algorithm \ref{enu:M3} calculates $U(x)$, $\sigma(x)$ and the Hessian matrix of
$U(x)$ once every iteration.

\begin{figure}[h]
  \centering
  \caption{Inversion of Random Coefficients Logit Demand}
  \vspace{0.5em}
  \includegraphics[width=0.9\columnwidth]{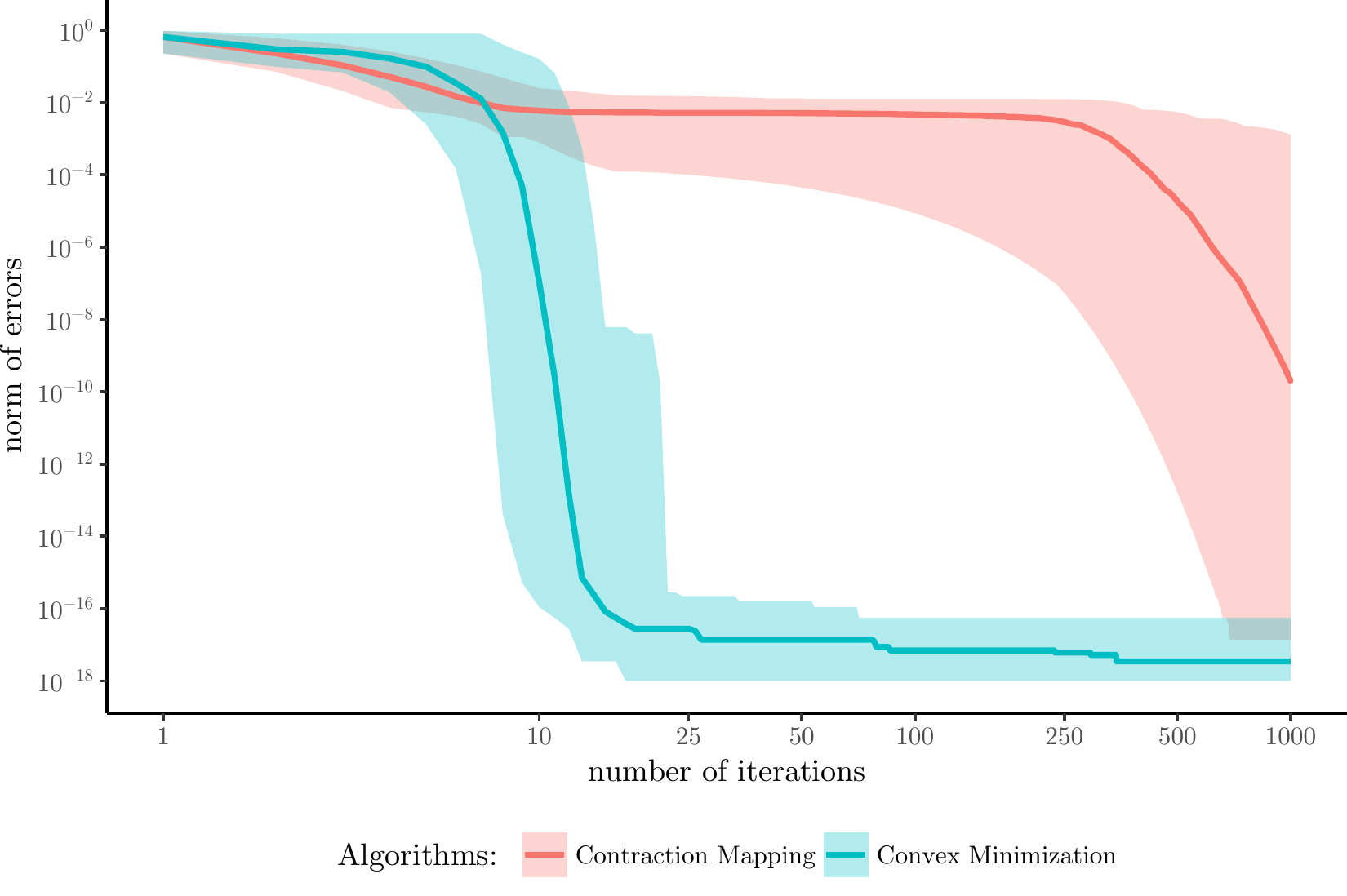}
  \label{fig:BLP}
\end{figure}

The result is shown in Figure \ref{fig:BLP}, where I plot the area between the
worst and the best norm of the error among all simulation experiments, and plot
the median of the norm of these errors in solid lines. The method based on
convex optimization clearly outperforms the contraction mapping algorithm. After
only $25$ iterations of the convex optimization method, the norm of error is
less than $10^{-15}$ in all simulation experiments. In contrast, after $250$
iterations of the contraction mapping algorithm, the norm of error is still
larger than $10^{-3}$ in more than half of the simulation experiments. Note the
error curve of the contraction mapping algorithm is almost flat during iteration
$10$ to $250$, indicating the Lipschitz constant of the algorithm, i.e. the
$\alpha$ in \eqref{eq:linear_contraction}, could be very close to $1$. This is
the major reason why the contraction mapping algorithm can be very slow. More
discussion on this can be founded in \cite{dube_improving_2012}.

\subsection{Pure Characteristics Models}
Let's construct a pure characteristics model as follows. Let $J$ denote the
number of products and $M$ denote the dimension of product attributes. For any
vector $y$ in $\real^M$, let $y^{(1)}$ denote its first coordinate and
$y^{(-1)}$ denotes its remaining $M-1$ coordinates.

In each simulation experiment, we normalize $\beta^{(1)} = 1$ and draw
$\beta^{(-1)}$ from a uniform distribution on $[0, 1]^{M-1}$. We then draw
$z_1,...,z_J$ independently from a normal distribution $N(0, I_{M})$. As for
random coeffcients $\nu$, we simulate $n= 5000$ i.i.d $\nu^{(-1)}_i$ samples
from $N(0, I_{M-1})$. Given $\beta$, $\{z_j\}$ and $\{\nu^{(-1)}_i\}$ in each
simulation, consumer welfare $U(x)$ and market share $\sigma(x)$ can be
calculated as
\begin{eqnarray*}
  U(x) & = & \frac{1}{n}\sum_{i = 1}^n \int u(x_i, z_i, \nu_i^{-1}, \nu^{(1)})
             \ud \varphi(\nu^{(1)}) \\
  \sigma_j(x) & = & \frac{1}{n}\sum_{i = 1}^n \int \indicator\left( x_j +
             z_j^{(-1)\prime}\nu_i^{(-1)} + z_j^{(1)}\nu^{(1)} \ge  u(x_i, z_i, \nu_i^{-1}, \nu^{(1)})\right)
             \ud \varphi(\nu^{(1)})
\end{eqnarray*}
where $\varphi(\cdot)$ denotes the density function of the standard normal
distribution $N(0,1)$, and
\begin{displaymath}
  u(x_i, z_i, \nu_i^{(-1)}, \nu_i^{(1)}) \coloneqq \max\left\{0, \max_j\left( x_j +
             z_j^{(-1)\prime}\nu_i^{(-1)} + z_j^{(1)}\nu^{(1)}\right) \right\}
\end{displaymath}
The, the market share $\sigma^*$ and its true inversion $x^*$ can be written as
\begin{eqnarray*}
  x^* & = & (z_1'\beta,...,z_J'\beta) \\
  \sigma^*  & = & \sigma(x^*)
\end{eqnarray*}

In the simulations, I solve the inversion problem $\sigma(x) = \sigma^*$ using the
following two methods:
\begin{enumerate}[label=(\alph*)]
\item\label{enu:M4} Trust-region algorithm applied directly to nonlinear
  equation $\sigma(x) = \sigma^*$.
\item\label{enu:M5} Trust-region algorithm applied to convex optimization
    \eqref{thm:convex_representation}.
\end{enumerate}
To implement Algorithm \ref{enu:M4}, I utilize Matlab function {\tt fsolve} with
its algorithm option set to {\tt trust-region}. When implementing Algorithm
\ref{enu:M5}, I use a self-implemented trust-region procedure. Algorithm
\ref{enu:M5} is guaranteed to converge no matter what the initial value is,
while Algorithm \ref{enu:M4} only ensures local converge when the initial value
is close enough to the true solution. Therefore, we mainly focus on comparing
the robustness of different methods.

Similar to the previous simulation experiments, I conduct $100$ simulations
experiments with $J = 10$ and $M = 5$. Within each simulation experiment, a new
set of $\beta^{(-1)}$, $\{z_j\}$ and $\{\nu_j^{(-1)}\}$ is drawn and each
algorithm takes $x_1 = x^* + \delta$ as its initial point, where $\delta$ is a
random vector drawn in each experiment with $\norm{\delta} = 20$. In each
iteration $k$, I record the maximum norm of the error, $\norm{\sigma(x^{(k)}) -
  \sigma^*}_{\infty}\coloneqq\max_j(|\sigma_j(x^{(k)}) - \sigma^*_j|)$, where
$x^{(k)}$ is the best solution up to iteration $k$. Algorithm \ref{enu:M4}
calculates $\sigma(x)$ and the Hessian matrix of $U(x)$ once every iteration,
and Algorithm \ref{enu:M5} calculates $U(x)$, $\sigma(x)$ and the Hessian matrix
of $U(x)$ once every iteration.

\begin{figure}[h]
  \centering
  \caption{Inversion of Pure Characteristics Demand}
  \includegraphics[width=0.9\columnwidth]{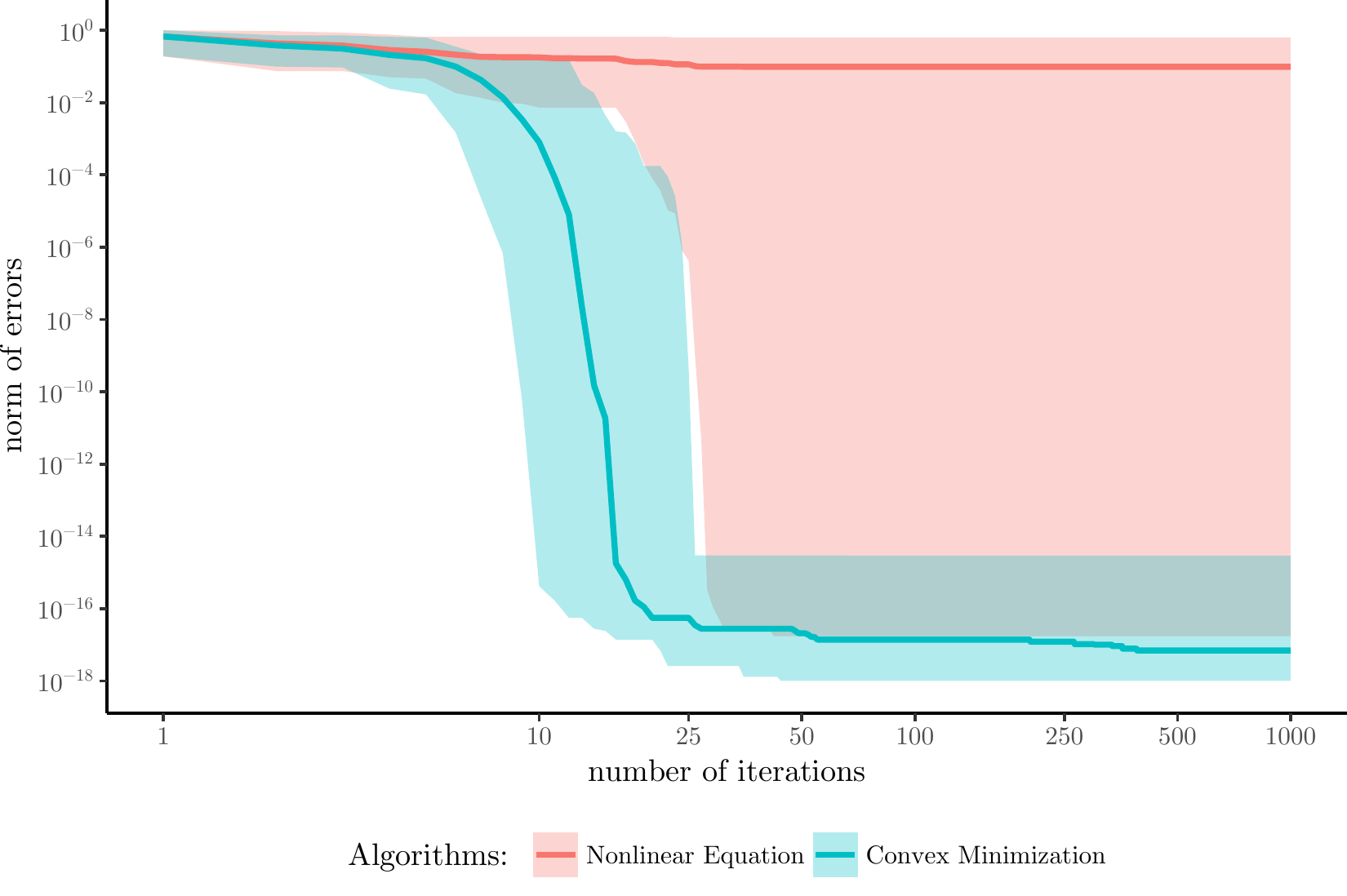}
  \label{fig:PureCh}
\end{figure}

The result is shown in Figure \ref{fig:PureCh}, where I plot the area between
the worst and the best norm of erros of all simulation experiments. I also plot
the median in solid lines. Clearly, the method based on convex minimization
dominates, in terms of both efficiency and robustness. After $25$ iterations of
the convex minimization method, the norm of errors is less than $10^{-14}$ in
all simulation experiments. In contrast, the method directly solving the
nonlinear equation $\sigma(x) = \sigma^*$ is much less robust. It fails to
converge in more than half of the experiments. And, when the norm of errors does
converges, it converges slower than the convex minimization method.

The performance difference between these two methods is mainly driven by
the singularity of the Hessian matrix. In more than half of our simulation
experiments, the minimum market share of products, $\min(\min_j\sigma^*_j, 1 -
\sum_j\sigma^*_j)$, is smaller than $10^{-14}$. In some of the experiments, the
minimum market share of products is even zero. In these cases, the Hessian matrix of
$U(x)$, or equivalently the Jacobian matrix of $\sigma(x)$, is near singular or
singular. As discussed in the end of Section \ref{sec:disc_choice}, the method
based on the convex optimization is able to utilize more information than the method
solving the nonlinear equation directly, which ensures its strong convergence
property even in those singular cases.

\newpage
\appendixpage
\addcontentsline{toc}{section}{Appendices}\markboth{APPENDICES}{}
\begin{appendices}
  \section{Proof of Theorem \ref{thm:convex_representation}}\label{sec:proof}
  First of all, by Theorem 23.5 in \cite{rockafellar_convex_1997}, in
  particular, by the equivalence of condition (a) and (b) in Theorem 23.5, we
  know condition \eqref{eq:convex_conjugate} is equivalent to $\sigma^* \in \partial
  U(x^*)$, where $\partial U(x)$ stands for the subgradient of $U(x)$ with respect
  to $x$.

  Let $y_i(x) \in \real^J$ be the choice of agent $i$, with $y_{i,j} = 1$ if product
  $j$ is chosen and $y_{i,j} = 0$ if not chosen. Then, 
  \begin{displaymath}
    y_i(x) \in \partial \tilde{U}(x, \epsilon_i)
  \end{displaymath}
  where $\tilde{U}(x,\epsilon_i) = \max\left(0, \max_{j=1,...,J}(x_j +
    \epsilon_{ij}) \right)$ is individual $i$'s utility with her random
  $\epsilon_i$, and $\partial \tilde{U}(x,\epsilon_i)$ stands for the
  subgradient of $\tilde{U}(x, \epsilon_i)$ with respect to $x$. With these
  notations, we can write average consumer welfare as $U(x) = \expt_F
  \tilde{U}(x, \epsilon_i)$ and market share as $\sigma(x) = \expt_F y_i(x)$.
  
  By Proposition 2.2 in \cite{bertsekas_stochastic_1973}, we know
  \begin{displaymath}
    \expt_F y_i(x) \in \partial\expt_{F} \tilde{U}(x,\epsilon_i), 
  \end{displaymath}
  or, equivalently, $\sigma(x) \in \partial U(x)$. Therefore, if $\sigma(x^*) =
  \sigma^*$, we must have $\sigma^* \in \partial U(x^*)$. Hence, we know
  condition \eqref{eq:convex_conjugate} must hold.

  If distribution $F$ is non-atomic, then $\partial\tilde{U}(x, \epsilon_i) =
  \{y_i\}$ almost surely. By proposition 2.2 in
  \cite{bertsekas_stochastic_1973}, this implies $\partial \expt_F
  \tilde{U}(x,\epsilon_i) = \{ \expt_F y_i\}$, or equivalently, $\partial U(x) =
  \{\sigma(x)\}$. When condition \eqref{eq:convex_conjugate} holds, we know
  $\sigma^* \in \partial U(x^*)$ and hence, $\sigma^* = \sigma(x^*)$.
\end{appendices}

\newpage
\bibliography{algorithms}


\end{document}